\documentclass[11pt]{article}
\usepackage{amssymb,amsmath,epsfig}
\def\one{{{{\rm 1} \kern -.19em {\rm l}}}}
\def\C{{{{\rm {\mbox{\small l}}} \kern -.50em {\rm C}}}}
\def\R{{{{\rm l} \kern -.15em {\rm R}}}}
\def\N{{{{\rm l} \kern -.15em {\rm N}}}}
\def\E{{{{\rm l} \kern -.15em {\rm E}}}}
\def\P{{{{\rm l} \kern -.15em {\rm P}}}}
\def\Z{{{{\rm Z} \kern -.35em {\rm Z}}}}
\def\1{{{{\rm 1} \kern -.35em {\rm 1}}}}

\begin{document}
\begin{sloppypar}
\vspace*{0cm}
\begin{center}
{\setlength{\baselineskip}{1.0cm}{ {\Large{\bf
PSEUDO-HERMITIAN AND \boldmath{${\cal PT}$}-SYMMETRIC QUANTUM SYSTEMS WITH ENERGY-DEPENDENT 
POTENTIALS: BOUND-STATE SOLUTIONS AND ENERGY SPECTRA \\}} }}
\vspace*{1.0cm}
{\large{\sc{Axel Schulze-Halberg}}}$^\dagger$ and {\large{\sc{Pinaki Roy}}}$^\ddagger$
\end{center}
\noindent \\

$\dagger$ Department of Mathematics and Actuarial Science and Department of Physics, Indiana University Northwest, 3400 Broadway,
Gary IN 46408, USA, e-mail: axgeschu@iun.edu, xbataxel@gmail.com \\ \\

$\ddagger$ Physics and Applied Mathematics Unit, Indian Statistical Institute, Kolkata 700108, India, 
e-mail: pinaki@isical.ac.in \\ \\

\vspace*{.5cm}
\begin{abstract}
\noindent
We introduce generalized versions of complex Scarf and Morse-type potentials that contain energy-dependent 
parameters. ${\cal PT}$-symmetry and pseudo-hermiticity of the associated quantum systems are discussed, and 
a modified orthogonality relation and pseudo-norm are constructed. We show that despite energy-dependence, 
our systems can admit real energy spectra and normalizable solutions of bound-state type.

\end{abstract}


\section{Introduction}
In recent years several interrelated approaches were developed that extend conventional Quantum Theory into the 
complex domain, where the governing Hamiltonian of a quantum system loses hermiticity. The principal interest in 
such non-hermitian Hamiltonians stems from the initial observation \cite{bender} that they can feature real spectra. 
A sufficient, but not necessary condition for real eigenvalues is ${\cal PT}$-symmetry of the underlying Hamiltonian, referring to 
invariance under parity and time-reversal. If the Hamiltonian has the standard form $H=p^2+V$, then ${\cal PT}$-symmetry 
amounts to the potential remaining the same if the simultaneous replacements 
$x \rightarrow -x$ and $i \rightarrow -i$ are made. There is a vast amount of literature on ${\cal PT}$-symmetric quantum systems and 
its applications. A comprehensive introduction to the topic can be found in \cite{cbintro} \cite{cbsense}, 
while for recent applications in optics, waveguides and electrodynamics the reader is referred to \cite{kalozoumis} 
\cite{zya} \cite{ruter} \cite{schindler} and references therein. Another concept that leads to a condition for non-hermitian 
Hamiltonians admitting real spectra is called pseudo-hermiticy \cite{mos}. This condition requires the existence of a 
certain similarity transformation mapping a non-hermitian Hamiltonian onto a counterpart that is hermitian in a 
$L^2$-Hilbert space. As a consequence of such a transformation, the non-hermitian Hamiltonian can be rendered hermitian 
in a weighted $L^2$ space. Similar to ${\cal PT}$-symmetry, pseudo-hermiticity is a sufficient, but not necessary condition for 
the reality of a Hamiltonian's spectrum. A complete discussion of pseudo-hermiticity and its relationship to related concepts like ${\cal PT}$-symmetry and 
quasi-hermiticity is provided in the monograph \cite{mos}. Recent applications of pseudo-hermitian quantum systems 
concern periodic Hamiltonians \cite{maa}, dynamical invariants \cite{lache}, relativistic models \cite{pinaki}, 
random matrix theory \cite{shashi}, among others. While the concepts of pseudo-hermiticity and ${\cal PT}$-symmetry have 
been studied extensively, they have not yet been applied to quantum systems that feature energy-dependent potentials 
\cite{formanek}. Applications of energy-dependent potentials can be found in magneto-hydrodynamic models of the dynamo
effect \cite{dynamo}, hydrodynamics \cite{hydro1}, the Hamiltonian formulation of relativistic
quantum mechanics \cite{rela2}, confined models \cite{lombard}, their generation by means of the 
supersymmetry formalism \cite{yekken} and through point transformations of hypergeometric equations \cite{jesus}, 
just to mention a few examples. Models with energy-dependent potentials require a modification of the 
underlying quantum theory that affects norm and completeness 
relation in a fundamental way \cite{hokkyo} \cite{hokkyoerr} \cite{miya}. The purpose of this work is to study 
several cases of complex, energy-dependent potentials that are associated with real energy spectra. In particular, 
we consider an energy-dependent version of the ${\cal PT}$-symmetric hyperbolic Scarf potential and its trigonometric 
counterpart. In the trigonometric case we are able to demonstrate how energy-dependence in the potential 
determines the number of bound states that the system admits. Our final example is an energy-dependent 
version of the pseudo-hermitian Morse-type potential. In all these cases, we show explicit formulas for the discrete energies, the associated 
solutions of bound-state type, and we establish both orthogonality and normalizability. The remainder of this paper 
is organized as follows: section 2 summarizes basic facts about quantum systems featuring pseudo-hermiticity or 
${\cal PT}$-symmetry. In section 3 we construct an orthogonality relation and pseudo-norm for systems with energy-dependent 
potentials. Sections 4 and 5 are devoted to the study of three particular quantum systems that have complex potentials 
dependent on the energy.

\section{Pseudo-hermiticity and \boldmath{${\cal PT}$}-symmetry}
We will now briefly introduce the concepts of pseudo-hermiticity and ${\cal PT}$-symmetry. For further information 
the reader is referred to the comprehensive monograph \cite{mos} and references therein. Now assume that a 
quantum system is governed by a Hamiltonian $H$, defined on a Hilbert space ${\cal H}$ or at least on a dense 
subset of that space. If there is a (self-adjoint), bounded automorphism $\eta$ on ${\cal H}$, such that its adjoint 
$H^\dagger$ satisfies $H^\dagger = \eta H \eta^{-1}$, then the Hamiltonian $H$ is said to be weakly 
($\eta$-) pseudo-hermitian. This property ensures that the Hamiltonian features a real spectrum, despite its 
potential being possibly complex-valued. Let us remark that for Hamiltonians of the standard form $H=p^2+V$, 
pseudo-hermiticity is established if the condition $V^\ast = \eta V \eta^{-1}$ is satisfied. 
While the concept of ${\cal PT}$-symmetry was shown to be a particular case of 
pseudo-hermiticity \cite{mos}, the explicit link between the two concepts is often hard to find. A ${\cal PT}$-symmetric quantum 
system is characterized by invariance of its Hamiltonian under parity and time-reversal. As mentioned in the 
introduction, for Hamiltonians of the standard form $H=p^2+V$, then ${\cal PT}$-symmetry 
amounts to the potential staying invariant under the simultaneous replacements 
$x \rightarrow -x$ and $i \rightarrow -i$. Before we conclude this brief review, a remark concerning 
quantum systems having energy-dependent potentials is in order. For these systems, the concepts of 
pseudo-hermiticity and ${\cal PT}$-symmetry cannot be applied in a rigorous way because the underlying quantum 
theory does not devise a Hilbert space as the domain of the Hamiltonian \cite{formanek}. More precisely, its domain is not 
equipped with an inner product and norm in the strict mathematical sense. We will discuss this property below in 
more detail.

\section{Orthogonality relation and pseudo-norm}
Since we will be appplying the concepts of pseudo-hermiticity and ${\cal PT}$-symmetry to systems with 
energy-dependent potentials, we need to define an appropriate orthogonality relation and norm for such systems. 
As is known from the conventional scenario involving real-valued, energy-dependent potentials \cite{formanek}, 
the continuity equation must be redefined, leading to a modified orthogonality relation which in turn determines the 
norm. We will now follow this process for the present case. Afterwards, we construct an alternative representation 
for the orthogonality relation that simplifies its evaluation. Let us further mention that detailed information on the topic 
can be found in \cite{formanek}, while the complex case is discussed in \cite{miya}, based on prior work 
\cite{hokkyo} \cite{hokkyoerr}.

\subsection{Construction of orthogonality relation and pseudo-norm}
In the first step, let us state the type of quantum systems we will focus on throughout this work. These systems are 
governed by a boundary-value problem of Dirichlet type. More precisely, let $D=(a,b)$ be an open interval on the 
real axis. We consider the problem
\begin{eqnarray}
\Psi''(x)+\left[E-V_E(x)\right] \Psi(x) &=& 0,~~~x \in D \label{bvp1} \\[1ex]
\Psi(a) ~=~ \Psi(b) &=& 0, \label{bvp2}
\end{eqnarray}
featuring pseudo-hermiticity or ${\cal PT}$-symmetric. As such, the potential $V_E$ of the 
governing Schr\"odinger equation is a complex function that is allowed to depend smoothly on the energy $E$. 
In order 
to construct solutions of (\ref{bvp1}), (\ref{bvp2}) that are physically meaningful, we have to first devise an appropriate 
orthogonality relation and norm for our system. Recall that if the potential is energy-dependent, we must resort to a 
modified definition of the orthogonality relation \cite{formanek}, given by
\begin{eqnarray}
\int\limits_a^b \left[1-\frac{V_{E_m}(x)-V_{E_n}(x)}{E_m-E_n}\right] \Psi_m^\ast(x)~\Psi_n(x)~dx &=& 
C~\delta_{mn}.
\end{eqnarray}
Here, $\Psi_m$ and $\Psi_n$ are solutions to our boundary-value problem (\ref{bvp1}), (\ref{bvp2}) at the 
respective energies $E_m$ and $E_n$, where $m \neq n$ are nonnegative integers. Furthermore, $C$ is a real-valued 
constant. Besides this modified definition, in our case 
we must also take pseudo-hermiticity or ${\cal PT}$-symmetry into account. To this end, we start from the continuity equation
\begin{eqnarray}
\frac{\partial P(x,t)}{\partial t} &=& - \frac{\partial J(x,t)}{\partial x}, \label{cont}
\end{eqnarray}
where $P$ and $J$ denote the probability density and the probability current, respectively. Let us for now assume that 
our system is pseudo-hermitian, such that the condition $\eta V_E \eta^{-1} = V^\ast_E$ is satisfied. The particular case of 
${\cal PT}$-symmetry will be dealt with afterwards. Now, since our system is 
pseudo-hermitian and has an energy-dependent potential, the continuity equation (\ref{cont}) must be modified 
\cite{formanek} \cite{mandal}. To this end, let us first define functions $\hat{\Psi}_j$, where $j$ is a nonnegative integer, as
\begin{eqnarray}
\hat{\Psi}_j(x,t) ~=~ \exp\left(-i ~E_j~ t \right) \Psi_j(x). \label{hatpsi}
\end{eqnarray}
Here, for each nonnegative integer $j$ this function is a solution to the time-dependent Schr\"odinger equation associated with (\ref{bvp1}), 
while we assume that $\Psi_j$ solves (\ref{bvp1}), (\ref{bvp2}) at the energy $E_j$, respectively. We can now 
rewrite the continuity equation (\ref{cont}) for the present case. Taking into account the definition (\ref{hatpsi}), we have
\begin{eqnarray}
\frac{\partial P(x,t)}{\partial t} +i\left[V^\ast_{E_n}(x)-V^\ast_{E_m}(x)\right] 
\hat{\Psi}_m^\ast(x,t)~\eta~\hat{\Psi}_n(x,t)  &=& -\frac{\partial J(x,t)}{\partial x}, \label{contmod}
\end{eqnarray}
where $m \neq n$. Recall that here we made the assumption $\eta V \eta^{-1} = V^\ast$. 
Now, a straightforward calculation shows that this modified continuity equation is satisfied by the following probability density $P$ 
and probability current $J$ 
\begin{eqnarray}
P(x,t) &=& \hat{\Psi}_m^\ast(x,t)~\eta~\hat{\Psi}_n(x,t) \label{prob} \\[1ex]
J(x,t) &=& -i \left[\hat{\Psi}_m^\ast(x,t)~\eta~\frac{\partial \hat{\Psi}_n(x,t)}{\partial x}-
\frac{\partial \hat{\Psi}_m^\ast(x,t)}{\partial x}~\eta~\hat{\Psi}_n(x,t)\right]. \label{curr}
\end{eqnarray}
Observe that (\ref{prob}) has the usual form of a probability density for pseudo-hermitian systems. 
In order to find the associated orthogonality relation, we integrate the left side of (\ref{contmod}) with 
respect to the variable $t$. Taking into account definitions (\ref{hatpsi}) and (\ref{prob}), this gives
\begin{eqnarray}
\int\limits^t  \hat{\Psi}_m^\ast(x,s)~\eta~\hat{\Psi}_n(x,s)+i\left[V^\ast_{E_n}(x)-V^\ast_{E_m}(x)\right] 
\hat{\Psi}_m^\ast(x,s)~\eta~\hat{\Psi}_n(x,s) ~ds ~=\nonumber \\[1ex]
& & \hspace{-7cm} =~ \left[1-\frac{V^\ast_{E_m}(x)-V^\ast_{E_n}(x)}{E_m-E_n}\right] 
\hat{\Psi}_m^\ast(x,t)~\eta~\hat{\Psi}_n(x,t). \label{intprob}
\end{eqnarray}
Now, integration of (\ref{intprob}) with respect to $x$ 
over the spatial domain $D=(a,b)$ gives the sought orthogonality relation 
\begin{eqnarray}
\int\limits_a^b \left[1-\frac{V^\ast_{E_m}(x)-V^\ast_{E_n}(x)}{E_m-E_n}\right] \Psi_m^\ast(x)~\eta~\Psi_n(x)~dx &=& 
C~\delta_{mn}, \label{ortho}
\end{eqnarray}
where $C$ is a real constant. Note that the time dependence has been discarded, as it does not affect integration 
over the spatial domain. Next, the associated pseudo-norm $N$ can be obtained by taking the limit $m \rightarrow n$, giving the 
result
\begin{eqnarray}
N\left(\Psi_n \right) &=& \int\limits_a^b \left[1-\frac{\partial V^\ast_{E_n}(x)}{\partial E_n}\right] \Psi_n^\ast(x)~\eta~\Psi_n(x)~dx. 
\label{norm}
\end{eqnarray}
Recall that both (\ref{ortho}) and (\ref{norm}) are only valid under the assumption 
$\eta V \eta = V^\ast$. In the presence of ${\cal PT}$-symmetry, the corresponding two identities can be constructed by a 
process similar to the one performed in this paragraph. We do not specify details of the calculation here, but 
merely state the result:
\begin{eqnarray}
& &\int\limits_a^b \left[1-\frac{V_{E_m}(x)-V_{E_n}(x)}{E_m-E_n}\right] \Psi_m^\ast(-x)~\Psi_n(x)~dx ~=~ 
C~\delta_{mn} \label{orthopt} \nonumber \\[1ex]
& & N\left(\Psi_n \right) ~=~ \int\limits_a^b \left[1-\frac{\partial V_{E_n}(x)}{\partial E_n}\right] \Psi_n^\ast(-x)~\Psi_n(x)~dx. 
\label{normpt}
\end{eqnarray}
If $N(\Psi_n)<\infty$ is real-valued for all admissible values of $n$, then the normalized solutions of our boundary-value 
problem (\ref{bvp1}), (\ref{bvp2}) are given by $\Psi_n/N(\Psi_n)$. Finally let us note that (\ref{norm}) and (\ref{normpt}) 
do not constitute a norm in the mathematical sense because they can become 
negative or take complex values.

\subsection{Wronskian representation for the orthogonality relation}
Let us now introduce a further tool that will prove useful in subsequent calculations. Since the solutions of our governing 
equation (\ref{bvp1}) are typically given in terms of special functions, exact resolution of the integral in 
our orthogonality relation (\ref{ortho}) is usually difficult. Using our continuity equation (\ref{contmod}), we will 
now show that the latter integral can be 
represented by means of Wronskians, such that their calculation is based entirely on 
derivatives without the need for integration. To this end, let us first focus on the case of pseudo-hermiticity, where the 
condition $\eta V_E \eta^{-1} = V^\ast_E$ applies. We make the following observation regarding 
our probability current (\ref{curr}) 
\begin{eqnarray}
i~J(x,t) &=& W_{\hat{\Psi}^\ast_m, \eta \hat{\Psi}_n}(x,t), \label{jw}
\end{eqnarray}
where $W$ denotes the Wronskian of the functions in its index, note that we adhere to our notation from the 
previous paragraph. Upon substitution of (\ref{jw}) into our continuity 
equation (\ref{contmod}) we get
\begin{eqnarray}
\frac{\partial}{\partial t} \left[ \hat{\Psi}_m^\ast(x,t)~\eta~\hat{\Psi}_n(x,t) \right] +i\left[V^\ast_{E_n}(x)-V^\ast_{E_m}(x)\right] 
\hat{\Psi}_m^\ast(x,t)~\eta~\hat{\Psi}_n(x,t)  = i~\frac{\partial}{\partial x} ~W_{\hat{\Psi}^\ast_m, \eta \hat{\Psi}_n}(x,t). 
\label{contw}
\end{eqnarray}
As mentioned before, here we made the assumption $\eta V \eta^{-1} = V^\ast$. In the ${\cal PT}$-symmetric case 
$\eta={\cal PT}$, the potential in (\ref{contw}) does not appear as a complex conjugate. Now, integration of (\ref{contw}) 
with respect to the variable $t$ by means of (\ref{hatpsi}) and afterwards 
cancelling all time-dependent terms and overall constant factors gives 
\begin{eqnarray}
\left[1-\frac{V^\ast_{E_m}(x)-V^\ast_{E_n}(x)}{E_m-E_n}\right] \Psi_m^\ast(x)~\eta~\Psi_n(x) &=& 
\frac{1}{E_n-E_m}~\frac{\partial}{\partial x} ~W_{\Psi^\ast_m, \eta \Psi_n}(x). \nonumber
\end{eqnarray}
Next, integration over the spatial domain $D=(a,b)$ results in
\begin{eqnarray}
\int\limits_a^b \left[1-\frac{V^\ast_{E_m}(x)-V^\ast_{E_n}(x)}{E_m-E_n}\right]  \Psi^\ast_m(x)~\eta~\Psi_n(x)~dx &=& 
\frac{W_{\Psi^\ast_m,\eta \Psi_n}(b)-W_{\Psi^\ast_m,\eta \Psi_n}(a)}{E_n-E_m}. \label{orthow}
\end{eqnarray}
We see that the left sides of this identity and of (\ref{ortho}) coincide. Consequently, 
we can now verify orthogonality in the sense (\ref{ortho}) by simply evaluating the Wronskian of the two involved 
solutions, without having to perform any integration. More precisely, we have
\begin{eqnarray}
W_{\Psi^\ast_m,\eta \Psi_n}(b)-W_{\Psi^\ast_m,\eta \Psi_n}(a) &=& K~\delta_{mn}, \nonumber
\end{eqnarray}
where $K$ is a real-valued constant. In particular, if the Wronskian vanishes at both ends of the domain, 
we have orthogonality of the functions $\Psi_m$ and $\Psi_n$. Before we conclude this section, let us state the 
identity (\ref{orthow}) in the case of ${\cal PT}$-symmetry. We have
\begin{eqnarray}
\int\limits_a^b \left[1-\frac{V_{E_m}(x)-V_{E_n}(x)}{E_m-E_n}\right]  \Psi^\ast_m(-x)~\Psi_n(x)~dx &=& 
\frac{W_{{\cal PT}\Psi_m,\Psi_n}(b)-W_{{\cal PT}\Psi_m,\Psi_n}(a)}{E_n-E_m}, \label{orthoptw}
\end{eqnarray}
note that ${\cal PT}\Psi_m(x) = \Psi^\ast(-x)$. As a direct consequence, we find
\begin{eqnarray}
W_{{\cal PT}\Psi_m,\Psi_n}(b)-W_{{\cal PT}\Psi_m,\Psi_n}(a) &=& K~\delta_{mn}, \nonumber
\end{eqnarray}
where as before $K$ is a real-valued constant. Again, if the Wronskian vanishes at both ends of the domain, 
we have orthogonality of the functions $\Psi_m$ and $\Psi_n$.

\section{Energy-dependent Scarf systems}
In this section we will focus on two models of Scarf type, the potentials of which are complex-valued and depend on the 
system's energy. Both of the potentials are non-hermitian, but can take a ${\cal PT}$-symmetric form if their 
parameters are chosen in a suitable way. 

\subsection{Hyperbolic Scarf system}
We consider the boundary-value problem (\ref{bvp1}), (\ref{bvp2}) for the domain $D=(-\infty,\infty)$. Note that in 
the present case the boundary conditions (\ref{bvp2}) are understood in the sense of limits. We introduce the following 
potential $V_E$ 
\begin{eqnarray}
V_E(x) &=& A^2+\left[B(E)^2-A^2-A~\alpha\right] \mbox{sech}^2(\alpha x)
+B(E)~(2 A+\alpha)~\tanh(\alpha x)~\mbox{sech}(\alpha x), \label{pot1}
\end{eqnarray}
where the right side contains two real-valued constants $A$, $\alpha>0$ and a complex-valued 
parameter $B$ that depends on $E$. We observe that (\ref{pot1}) is a complex, energy-dependent generalization of the 
known complex hyperbolic Scarf potential. An inspection reveals that (\ref{pot1}) is not hermitian due to the 
complex parameters it contains. It is of interest to note that $\cal{PT}$-symmetric \cite{z1} as well as the non 
$\cal{PT}$-symmetric \cite{zafarx} versions of this potential possess a real-valued 
discrete spectrum. In particular, the potential (\ref{pot1}) 
becomes the conventional ${\cal PT}$-symmetric hyperbolic Scarf potential \cite{z1} if the parameter $B$ is purely imaginary. This can be easily verified by 
applying the simultaneous replacements $x \rightarrow -x$, $i \rightarrow -i$ and using symmetry properties of 
the hyperbolic functions. It is known that our system with potential (\ref{pot1}) must fulfill a particular condition 
in order for ${\cal PT}$-symmetry to be unbroken \cite{ahmedbr}, we will evaluate this condition below when 
going through an example. Next, let us state the solutions of 
our boundary-value problem (\ref{bvp1}), (\ref{bvp2}) for the potential (\ref{pot1}). 
These solutions exist if the energies take 
discrete values $E=E_n$,  where $n$ is an integer satisfying $0 \leq n < A/\alpha$, given by
\begin{eqnarray}
E_n &=& A^2-(A-\alpha~n)^2. \label{ene1}
\end{eqnarray}
Note that the first term $A^2$ on the right side has been introduced here and in (\ref{pot1}) in order to 
fix the lowest energy at zero, that is, we have $E_0=0$. Next, observe that the energies (\ref{ene1}) 
are real-valued, even though the potential (\ref{pot1}) takes complex values. This 
gives rise to the conjecture of $V_E$ being $\cal{PT}$-symmetric, as mentioned above. The solutions 
$\Psi=\Psi_n$, $0 \leq n < A/\alpha$, of our equation (\ref{bvp1}) associated with the discrete energies (\ref{ene1}), are given by
\begin{eqnarray}
\Psi_n(x) &=& \exp\left\{
-\frac{B(E_n)}{\alpha} ~\arctan\left[\sinh\left(\alpha x\right) \right]\right\}
\mbox{sech}^{\frac{A}{\alpha}}\left(\alpha x \right) \times \nonumber \\[1ex]
& & \hspace{3cm} \times ~P_n^{\left(-\frac{1}{2}-\frac{A}{\alpha}+i \frac{B(E_n)}{\alpha},
-\frac{1}{2}-\frac{A}{\alpha}-i \frac{B(E_n)}{\alpha}\right)}\left[
-i \sinh\left(\alpha x \right)\right], \label{sol1}
\end{eqnarray}
where $P_n$ stands for the Jacobi polynomial of degree $n$ \cite{abram}. Here we are making the assumption that our parameters 
$A$, $B$ and $\alpha$ are chosen such that the values $B(E_n)$ in (\ref{sol1}) remain defined for all $0 \leq n < A/\alpha$. 
Let us now briefly discuss the asymptotic behaviour of our solutions 
(\ref{sol1}) at the infinities. To this end, we first observe that both real and imaginary part of the exponential term remain 
bounded on the real line, independent of the values that $B(E_n)$ and $\alpha$ attain. As a 
consequence, the solutions behave at the infinities like $\Psi_n(x) \sim \sinh^n(\alpha x)~ \mbox{sech}^{A/\alpha}(\alpha x)$ in 
real and imaginary part. Since we are requiring $n<A/\alpha$, it follows $\Psi_n(x) \rightarrow 0$ for $|x| \rightarrow \infty$, 
such that our boundary conditions (\ref{bvp2}) are satisfied if $a$ and $b$ are chosen as negative and positive 
infinity, respectively. It remains to evaluate the orthogonality relation (\ref{ortho}) and the pseudo-norm (\ref{norm}). Since the 
resulting expressions are very large if the general form (\ref{sol1}) of the solutions is used, we will restrict ourselves 
to particular examples. 

\begin{figure}[h]
\begin{center} \vspace{.3cm}
\epsfig{file=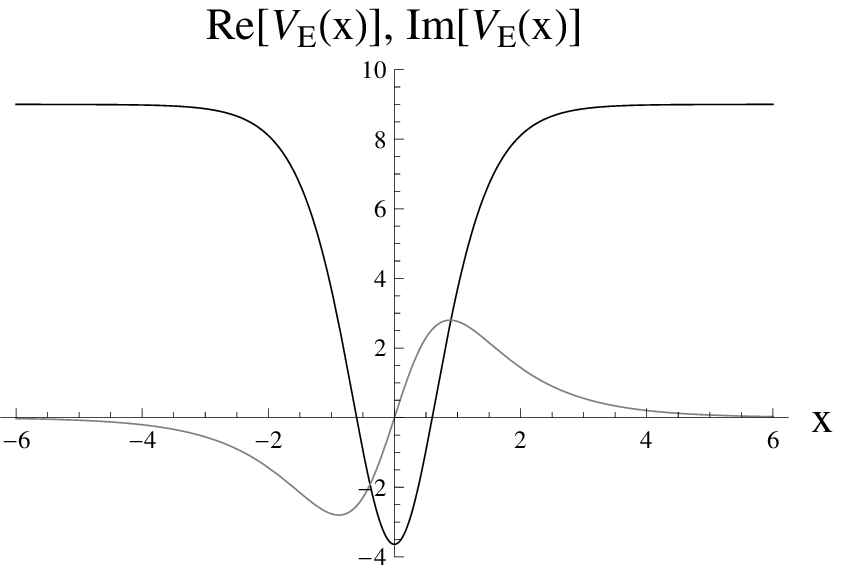,width=7.8cm}
\epsfig{file=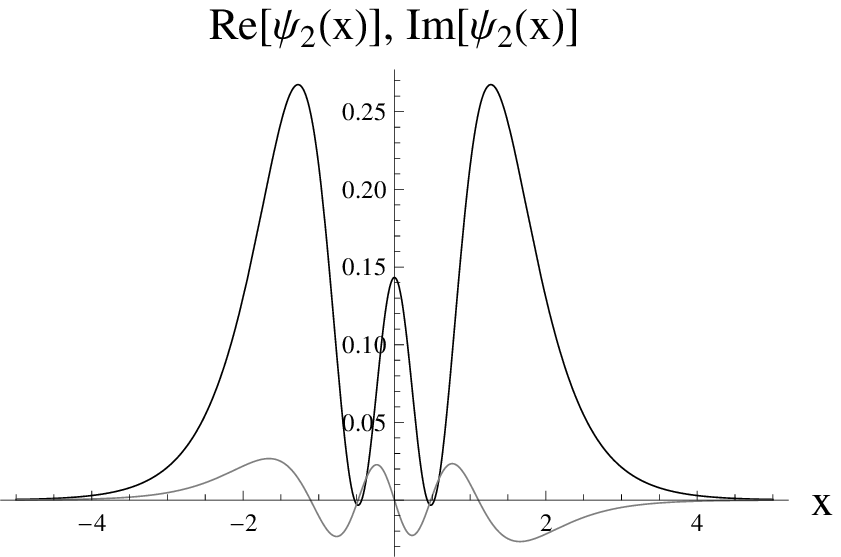,width=7.8cm}
\caption{Left plot: graph of the real part (black curve) and imaginary part (grey curve) of the potential (\ref{pot1ptx}) 
with $n=2$ and $k=0.1$. Right plot: graph of the real part (black curve) and imaginary part (grey curve) 
of (\ref{norm1x}) for $n=2$ and $k=0.01$. }
\label{fig1}
\end{center}
\end{figure}

\paragraph{Example:  \boldmath{${\cal PT}$}-symmetric case.} Let us now pick particular values for the parameters that determine the potential 
(\ref{pot1}). We set 
\begin{eqnarray}
A~=~3 \qquad \qquad \qquad \alpha~=~1 \qquad \qquad \qquad B(E)&=& i~k~E, \label{set1}
\end{eqnarray}
where $k>0$ is a constant. Before we continue, let us briefly justify these parameter settings. The values of $A$ 
and $\alpha$ must be such that the system admits bound-state solutions. Since $\alpha$ provides merely a 
scaling of the variable, we can without restriction set it equal to one. Furthermore, we pick an integer value for $A$ 
in order to simplify calculations and notation. For the same reason we choose a $B$ that depends linearly on the 
energy. We point out that other dependencies may be chosen, such as exponential or rational. 
Now, since $B$ is purely imaginary, the potential (\ref{pot1}) becomes ${\cal PT}$-symmetric. After incorporation of (\ref{set1}), 
it takes the form 
\begin{eqnarray}
V_E(x) &=& 9+[-12-k^2~n^2~(n-6)^2] ~\mbox{sech}^2(x)-7~i~k~n(n-6)~ \mbox{sech}(x)~\tanh(x). \nonumber
\end{eqnarray}
\vspace{-1.5cm} \\
\begin{eqnarray}
\label{pot1ptx}
\end{eqnarray}
The left plot of figure \ref{fig1} shows a graph of this potential. Next, the discrete energies of our boundary-value problem (\ref{bvp1}), (\ref{bvp2}) can now be found 
by substituting the settings (\ref{set1}) into (\ref{ene1}). This yields 
\begin{eqnarray}
E_n &=& n~(6-n),~~~0 \leq n < 3. \label{ene1x}
\end{eqnarray}
Before we state the associated solutions of our system, let us ensure that ${\cal PT}$-symmetry is unbroken. 
To this end, the following condition must be satisfied \cite{ahmedbr}
\begin{eqnarray}
-7~k~n~(n-6) &<& \left[k~n~(n-6)\right]^2+\frac{49}{4}. \label{ineq}
\end{eqnarray}
Since $n$ can only take the values $0,1$ and $2$, we will plug each of these into (\ref{ineq}) in order to find out 
which values of $k$ are suitable choices. For $n=0$ the inequality (\ref{ineq}) is always satisfied without any restriction 
on $k$. Upon setting $n=1$ we find that (\ref{ineq}) is fulfilled if $k < 7/10$. Finally, in case $n=2$, condition 
(\ref{ineq}) is satisfied for all values of $k$ except $k=7/16$. Consequently, in order to guarantee that 
${\cal PT}$-symmetry is unbroken in our system, we can choose any positive value for $k$ that is less than $7/10$ and 
different from $7/16$. Next, the solutions $\Psi_n$, $0 \leq n < 3$, of our equation (\ref{bvp1}) 
are obtained from (\ref{sol1}) by plugging in the settings (\ref{set1}). These solutions read as follows
\begin{eqnarray}
\Psi_n(x) &=& \exp\big\{-i~k~n~(n-6)~ \arctan[\sinh(x)]\big\} ~\mbox{sech}^3(x) ~\times \nonumber \\[1ex]
& & \hspace{3cm} \times~
P_n^{\left(kn(n-6)-\frac{7}{2},-kn(n-6)-\frac{7}{2} 
\right)}\left[-i \sinh(x)\right], \label{sol1x}
\end{eqnarray}
where $0 \leq n <3$. Let us now show orthogonality of the functions in (\ref{sol1x}) with respect to the relation 
(\ref{ortho}). Since we are unable to evaluate the integral contained in the latter relation, we use its Wronskian 
representation (\ref{orthow}). For admissible indices $m$, $n$ satisfying $m \neq n$, 
we find the Wronskian of ${\cal PT}\Psi_m$ and $\Psi_n$ as
\begin{eqnarray}
W_{{\cal PT}\Psi_m,\Psi_n}(x) &=& \nonumber \\[1ex] 
& & \hspace{-3cm} =~\frac{i}{2}~\exp\Bigg\{ \Big[i~k~\left(m^2-6~m+n^2-6~n \right)\Big]
\arctan[\sinh(x)]\Bigg\}~\mbox{sech}^7(x)
 \times \nonumber \\[1ex]
& & \hspace{-3cm} \times~\Bigg\{(m-6)~\cosh^2(x)~P^\ast_{m-1}(x)~P_n(x)+2~P^\ast_m(x)~P_n(x)
~\Big[-k~m~(m-6)-3~i~\sinh(x)\Big]+ \nonumber \\[1ex]
& & \hspace{-3cm} + ~
(6-n)~\cosh^2(x)~P^\ast_{m}(x)~P_{n-1}(x)+2~P^\ast_m(x)~P_n(x)~P_n(x)
~\Big[k~n~(n-6)+3~i~\sinh(x)\Big] \Bigg\}
. \nonumber \\[1ex] \label{wron1}
\end{eqnarray}
Note that for the sake of brevity we used the following abbreviation for the Jacobi polynomials:
\begin{eqnarray}
P_{j-l}(x) &=& \left\{
\begin{array}{lll}
P_{j-l}^{\left(l+kj(j-6)-\frac{7}{2},l-kj(j-6)-\frac{7}{2} 
\right)}\left[-i \sinh(x)\right] & \mbox{if} & j-l \geq 0\\[1ex]
0 & \mbox{if} & j-l < 0
\end{array}
\right\}, \label{pjl}
\end{eqnarray}
where $j,l$ are nonnegative integers for $0 \leq j < 3$ and $0 \leq l \leq 1$. Now, inspection of (\ref{wron1}) shows that 
its first (exponential) term on the right side stays bounded in both real and imaginary part. Furthermore, 
the remaining terms tend to zero as $x$ goes to positive or negative infinity, since the power of the hyperbolic secant 
is larger than the power of the hyperbolic functions inside the brackets. As a consequence, real and imaginary part of 
the Wronskian (\ref{wron1}) vanish at positive and negative infinity. Therefore, we have in particular
\begin{eqnarray}
\lim\limits_{x \rightarrow \infty} W_{{\cal PT}\Psi_m,\Psi_n}(x) - \lim\limits_{x \rightarrow -\infty} W_{{\cal PT}\Psi_m,\Psi_n}(x) &=& 0. \nonumber
\end{eqnarray}
Upon plugging this into the left side of (\ref{orthow}), we obtain orthogonality of the functions in (\ref{sol1x}). 
Let us now proceed to establish normalizability of those functions by verifying that (\ref{normpt}) is finite. Since 
we are not able to resolve the integral on the right side of (\ref{normpt}) in closed form, we will proceed differently. 
In the first step, let us show that 
the integral in (\ref{normpt}) exists for the present case. To this end, we will now calculate 
the integrand in (\ref{normpt}). Upon substitution of (\ref{pot1ptx}), (\ref{ene1x}) and (\ref{sol1x}), we obtain
\begin{eqnarray}
\Psi^\ast_n(-x)~\Psi_n(x)\left[1-\frac{\partial V_{E}(x)}{\partial E}\right]_{\big| E=E_n} \hspace{-.4cm} &=& 
\frac{1}{2}~
\exp\Bigg\{2~i~k~n~(n-6) \arctan[\sinh(x)]\Bigg\} \times \nonumber \\[1ex]
& & \hspace{-6cm} \times~\mbox{sech}^8(x)~\Bigg\{1-4~k^2~n~(n-6)+\cosh(2~x)-14~i~k~\sinh(x) \Bigg\}~P^\ast_n(x)~P_n(x),
\label{norm1x}
\end{eqnarray}
where $0 \leq n <3$ and we have used the abbreviation (\ref{pjl}). The right plot in figure \ref{fig1} shows a particular case of (\ref{norm1x}). 
Similar to the case of the Wronskian (\ref{wron1}), the exponential terms in (\ref{norm1x}) remain 
bounded, such that behaviour at the infinities is determined by the hyperbolic secant function that is taken to the 
eighth power. Since this function falls 
more rapidly than the remaining hyperbolic sine and cosine functions grow, the function (\ref{norm1x}) 
tends exponentially to zero at both positive and negative infinify. As a consequence, the integral in (\ref{norm}) exists. 
Now, since there are only three functions in (\ref{sol1x}) due to the constraint $n<3$, we can calculate their norm integrals 
numerically once a particular value for the parameter $a$ is chosen. We pick $k=0.1$, this gives
\begin{eqnarray}
\begin{array}{llllllll}
N(\Psi_0) ~=~1.06667 & & N(\Psi_1) ~=~ -1.66462 & & N(\Psi_2) ~=~ 0.745127 . 
\end{array}  \label{norm1res}
\end{eqnarray}
We observe that our pseudo-norm alternates in sign. This is a known phenomenon to occur in the context of 
${\cal PT}$-symmetry \cite{levai}. Since going into further details regarding alternating-sign pseudo-norms is beyond the 
scope of this work, we refer the reader to \cite{cbintro}. Now, using the results from (\ref{norm1res}), we are now able to 
normalize our solutions (\ref{sol1x}). The correctly normalized solutions are 
simply given by $\hat{\Psi}_n=\Psi_n / N(\Psi_n)$, $0 \leq n <3$.

\paragraph{Example: generalization of Yekken's system.} In this paragraph we will show that our boundary-value 
problem (\ref{bvp1}), (\ref{bvp2}) with potential (\ref{pot1}) generalizes a model studied by 
Yekken and collaborators \cite{lombsing}. In the latter reference, the following potential is considered
\begin{eqnarray}
V_E(x) &=& -\lambda~(1+\gamma~E_n^\nu)~\mbox{sech}^2(x), \label{lom}
\end{eqnarray}
where $\lambda$, $\nu$ and $\gamma$ are real-valued constants. In order to recover a generalization of (\ref{lom}), 
we apply the following settings to our potential (\ref{pot1})
\begin{eqnarray}
A~=~-\frac{\alpha}{2}+\frac{1}{2}~\sqrt{\alpha^2+4~\lambda} \qquad \qquad 
B(E)~=~i~\exp\left(\frac{\nu}{2}\right)~\sqrt{\gamma~\lambda}. \label{setlom}
\end{eqnarray}
Upon substitution of these parameters, the potential (\ref{pot1}) takes the form
\begin{eqnarray}
V_E(x) &=& \frac{1}{4}\left(-\alpha+\sqrt{\alpha^2+4~\lambda}\right)-
\lambda~\left(1+\gamma~E_n^\nu\right) \mbox{sech}^2\left(\alpha x \right)+ \nonumber \\[1ex]
&+& i~E_n^{\frac{\nu}{2}} \sqrt{\gamma~\lambda~(\alpha^2+4~\lambda)}~\mbox{sech}(\alpha x)~
\tanh(\alpha x), \label{lomgen}
\end{eqnarray}
recall that the energies $E_n$ are given in (\ref{ene1}). 
Since for $\alpha=1$ the second term on the right side coincides with its counterpart in (\ref{lom}), we have 
constructed a complex, energy-dependent generalization of the latter potential. The discrete energies and 
associated solutions to the boundary-value problem with potential (\ref{lomgen}) can be obtained by 
plugging the settings (\ref{setlom}) into (\ref{ene1}) and (\ref{sol1}), respectively. For the sake of brevity we 
omit to state the resulting expressions.

\subsection{Trigonometric Scarf system}
The next application we present is governed by the boundary-value problem 
(\ref{bvp1}), (\ref{bvp2}), defined on the bounded domain $D=(-\pi/(2 \alpha),\pi/(2 \alpha))$ for a constant 
$\alpha>0$ and endowed with the potential
\begin{eqnarray}
V_E(x) &=&  -A^2+\left[B(E)^2+A^2-A~\alpha\right] \mbox{sec}^2(\alpha x)
-B(E)~(2 A-\alpha)~\tan(\alpha x)~\mbox{sec}(\alpha x), \label{pot3}
\end{eqnarray}
where the constants $A$, $\alpha$ are positive and $B$ is a complex-valued function of the energy $E$. We 
observe that (\ref{pot3}) generalizes the trigonometric Scarf potential, 
featuring energy-dependence and complex values. It should be mentioned that (\ref{pot3}) can be 
generated from its hyperbolic counterpart (\ref{pot1}) by making the simultaneous replacements 
$A \rightarrow -A$, $B \rightarrow B^\ast$, $x \rightarrow i x$ and flipping the overall sign of the potential. 
Since the discrete energies (\ref{ene1}) in the hyperbolic case do not depend on the parameter $B$, the 
aforementioned link between the hyperbolic and the trigonometric case implies that our potential (\ref{pot3}) is associated 
with real discrete energies and that it becomes ${\cal PT}$-symmetric if the real part of $B$ vanishes. 
Solutions of our boundary-value 
problem (\ref{bvp1}), (\ref{bvp2}) for the potential (\ref{pot3}) can be constructed provided the energy 
takes discrete values $E=E_n$, where $n$ is a nonnegative integer. These energy values are given by
\begin{eqnarray}
E_n &=& -A^2+(A+\alpha~n)^2,~~~n=0,1,2,... \label{ene3}
\end{eqnarray}
As in the previous two applications, these discrete energies are real-valued. Now, the solutions of our problem 
associated with (\ref{ene3}) are given by 
\begin{eqnarray}
\Psi_n(x) = \left[1-\sin\left(\alpha x \right)\right]^{\frac{A-B(E_n)}{2 \alpha}} 
\left[1+\sin\left(\alpha x \right)\right]^{\frac{A+B(E_n)}{2 \alpha}}
 P_n^{\left( -\frac{1}{2}+\frac{A}{\alpha}-\frac{B(E_n)}{\alpha},-\frac{1}{2}+\frac{A}{\alpha}+\frac{B(E_n)}{\alpha}
\right)}\left[\sin\left(\alpha x \right)\right]. \label{sol3}
\end{eqnarray}
As before, $P_n$ denotes the Jacobi polynomial of degree $n$. In order to check that our boundary conditions 
(\ref{bvp2}) are fulfilled at $\pm ~\pi/(2 \alpha)$, we inspect the first two factors on the right side of 
(\ref{sol3}). These factors vanish at $-\pi/(2 \alpha)$ and $\pi/(2 \alpha)$, respectively, provided 
the real part of their exponents remains positive. This is true as long as the constraint
\begin{eqnarray}
|\mbox{Re}[B(E_n)]| &<& A, \label{constraint}
\end{eqnarray}
is satisfied. This constraint leads to an interesting situation concerning the number of solutions (\ref{sol3}) to our 
boundary-value problem. Since $E_n$ changes with $n$, so do $B(E_n)$ and the functions (\ref{sol3}). 
It is therefore possible that for certain 
values of $n$ the constraint (\ref{constraint}) is not fulfilled, such that the associated solutions from (\ref{sol2}) 
do not satisfy the boundary conditions. As an illustration let us now assume that $A=5, \alpha=1$ and 
Re$[B(E_n)]=1/10$. The constraint (\ref{constraint}) then takes the form 
\begin{eqnarray}
\frac{n~(n+10)}{10} &<& 5. \nonumber
\end{eqnarray}
Solving for $n$ gives the result
\begin{eqnarray}
n ~<~-5+5~\sqrt{3} ~\approx~ 3.66025. \label{nest}
\end{eqnarray}
Since $n$ takes only integer values, the latter identity amounts to $n\leq3$. Consequently, our boundary-value problem 
(\ref{bvp1}), (\ref{bvp2}) for the trigonometric Scarf potential (\ref{pot3x}) has four solutions (\ref{sol3x}), 
corresponding to the parameter values $n=0,...,3$. Since in this example the real part of $B$ does not vanish, the 
underlying system is not ${\cal PT}$-symmetric.

\begin{figure}[h]
\vspace{.3cm}
\begin{center}
\epsfig{file=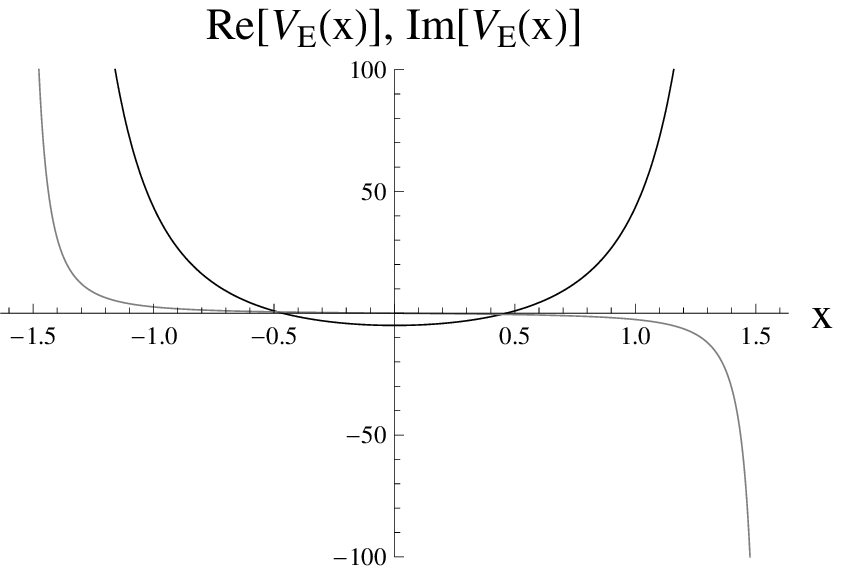,width=7.8cm}
\epsfig{file=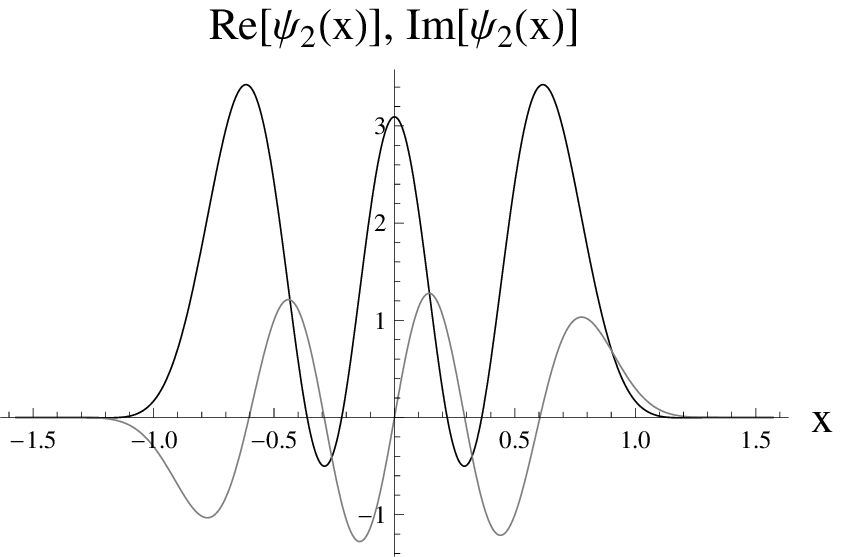,width=7.8cm}
\caption{Left plot: graph of the real part (black curve) and imaginary part (grey curve) of the potential (\ref{pot3x}) 
with $n=0$, $b=0.1$. Right plot: graph of the real part (black curve) and imaginary part (grey curve) of 
(\ref{dens3}) for $n=2$, $b=0.01$.}
\label{fig3}
\end{center}
\end{figure}

\paragraph{Example.} Let us now choose the parameters in our potential (\ref{pot3}) as follows
\begin{eqnarray}
A~=~5 \qquad \qquad \qquad \alpha~=~1 \qquad \qquad \qquad B(E)~=~b~\sqrt{E+1}, \label{set3}
\end{eqnarray}
where $b>0$ is a constant. Our choices for the parameter values of $A$ and $\alpha$ follow the same 
reason as (\ref{set1}) in the previous example. Similarly, $B$ was chosen to have a simple energy-dependence. Note 
that we must add one inside the radicand, as otherwise the lowest bound state solution for $n=0$ will be undefined. 
Since the parameter $B$ is purely imaginary, the system governed by the potential (\ref{pot3}) is 
${\cal PT}$-symmetric. In addition, the limiting condition (\ref{constraint}) is always satisfied. Upon 
substitution of our settings (\ref{set3}), the latter potential takes the form
\begin{eqnarray}
V_E(x) = -25+\Bigg\{20-b^2 \Bigg[1+n~(n+10) \Bigg] \Bigg\} \sec^2(x)
-  9~i~b ~\sqrt{1+n~(n+10)} \Bigg] \sec(x) \tan(x). \nonumber \\[1ex] \label{pot3x}
\end{eqnarray} 
A particular case of this potential is shown in the left plot of figure \ref{fig3}. 
Our boundary-value problem (\ref{bvp1}), (\ref{bvp2}) for the potential (\ref{pot3x}) admits solutions if the 
energy takes the discrete values obtained from (\ref{ene3}) after substitution of our settings (\ref{set3})
\begin{eqnarray}
E_n &=& n~(n+10),~~~n=0,1,2,... . \label{ene3x}
\end{eqnarray}
The solutions associated with these energies can be constructed by means of plugging (\ref{set3}) into 
(\ref{sol3}). This yields
\begin{eqnarray}
\Psi_n(x) &=&  \left[1-\sin\left(x \right)\right]^{
\frac{5}{2}-\frac{i b \sqrt{n~(n+10)+1}}{2}} 
\left[1+\sin\left(x \right)\right]^{
\frac{5}{2}+\frac{i b \sqrt{n~(n+10)+1}}{2}}  \times \nonumber \\[1ex]
&\times&
P_n^{\left( 
\frac{9}{2}-i b \sqrt{n(n+10)+1},\frac{9}{2}+i b \sqrt{n(n+10)+1}
\right)}\left[\sin\left(x \right)\right], \label{sol3x}
\end{eqnarray}
where $n=0,1,2,...$. Next, let us establish orthogonality of our solutions using the 
Wronskian representation (\ref{orthow}). We get for admissible integers $m,n$ with $m \neq n$
\begin{eqnarray}
W_{{\cal PT}\Psi_m,\Psi_n}(x) &=& \frac{1}{2} \left[\frac{1-\sin\left(x \right)}{1+\sin\left(x \right)}\right]^
{-\frac{1}{2}ib \left[\sqrt{m(m+10)+1}+\sqrt{n(n+10)+1}\right]
} ~\cos^9(x)~\times \nonumber \\[1ex]
& & \hspace{-2cm} \times~
~\Bigg\{(m+10)~\cos^2(x)~P^\ast_{m-1}(-x)~P_n(x)+2~P^\ast_m(-x)~P_n(x) 
\Bigg[(n+1))~\cos^2(x)~P_{n-1}(x)- \nonumber \\[1ex]
& & \hspace{-2cm}
-~2~i~b~\left[\sqrt{m(m+10)+1}+\sqrt{n(n+10)+1} \right] P_n(x)
\Bigg]
\Bigg\}
, \label{wron3}
\end{eqnarray}
where the following abbreviation is in use
\begin{eqnarray}
P_{j-k}(x) = \left\{
\begin{array}{lll}
 P_{j-k}^{\left( 
k+\frac{9}{2}-i b \sqrt{j(j+10)+1},k+\frac{9}{2}+i b \sqrt{j(j+10)+1}
\right)}\left[\sin\left(x \right)\right]
 & \mbox{if} & j-k \geq 0\\[1ex]
0 & \mbox{if} & j-k < 0
\end{array}
\right\}. \nonumber
\end{eqnarray}
\vspace{-0.6cm}
\begin{eqnarray}
\label{abb3}
\end{eqnarray}
Note that $j,k$ are nonnegative integers. It is straightforward to verify that 
the first two terms on the right side of (\ref{wron3}) guarantee vanishing of the Wronskian at $\pm \pi/2$ in 
both real and imaginary part. As an immediate consequence the left side of (\ref{orthoptw}) equals zero upon 
identifying $a=-\pi/2$ and $b=\pi/2$. This implies that the functions in (\ref{sol3x}) are orthogonal 
with respect to (\ref{orthopt}). Now, the remaining task is to evaluate the pseudo-norm (\ref{normpt}) for our solutions. After 
substituting (\ref{pot3x}), (\ref{sol3x}) and (\ref{ene3x}), the integrand on the right side of (\ref{normpt}) takes the following form
\begin{eqnarray}
\Psi^\ast_n(-x)~\Psi_n(x) \left[1-\frac{\partial V_{E}(x)}{\partial E}\right]_{\big| E=E_n} \hspace{-.5cm} &=& \nonumber \\[1ex]
& & \hspace{-4cm}=~
\left[1-\sin\left(x \right)\right]^{
5-ib\sqrt{n(n+10)+1}
} 
\left[1+\sin\left(x \right)\right]^{
5+ib\sqrt{n(n+10)+1}
} \times \nonumber \\[1ex]
& & \hspace{-4cm} \times~P^\ast_n(-x)~P_n(x)
\Bigg[
1+b^2~\sec^2(x)+\frac{9~i~b}{2~\sqrt{n~(n+10)+1}}~\sec(x)~\tan(x)
\Bigg]
. \label{dens3}
\end{eqnarray}
Observe that we have used the abbreviation (\ref{abb3}). The function (\ref{dens3}) is displayed in the right plot of 
figure \ref{fig3} for a particular parameter setting.
Similar to the case of (\ref{wron3}), we can show that both 
real and imaginary part of the 
first two terms on the right side of (\ref{dens3}) vanish at $\pi/2$ and $-\pi/2$, respectively. Continuity of 
(\ref{dens3}) then implies existence of the pseudo-norm integral in (\ref{norm}). Since the constraint (\ref{constraint}) is 
satisfied, there is no restriction on the number of bound-state solutions conatined in (\ref{sol3x}). Let us now 
determine their first four pseudo-norm integrals (\ref{normpt}) numerically after assigning a specific value to our parameter $b$. 
Upon setting $b=0.01$, we obtain
\begin{eqnarray}
N(\Psi_0) &=& 0.773111 \nonumber \\[1ex]
N(\Psi_1) &=& -1.94853 \nonumber \\[1ex]
N(\Psi_2) &=& 3.20699 \nonumber \\[1ex]
N(\Psi_3) &=& -4.38392. \nonumber
\end{eqnarray}
As in section 4.1, the pseudo-norm is alternating in sign \cite{cbintro} \cite{levai}. Upon using the above 
values, the correctly normalized solutions in (\ref{sol3x}) are obtained as $\Psi_n/N(\Psi_n)$, 
$n=0,1,2,...$.

\section{Energy-dependent Morse-type system}
In our next application our boundary-value problem is defined on the domain $D=(-\infty,\infty)$ and equipped with 
the following potential
\begin{eqnarray}
V_E(x) &=& A^2+B(E)^2~\exp\left(-2~\alpha~x \right)-2~B(E)~\left(A+\frac{\alpha}{2} \right)~\exp\left(-\alpha~x \right), \label{pot2}
\end{eqnarray}
introducing positive constants $A$ and $\alpha$ are positive, while $B$ is a complex-valued function of the energy $E$. 
The function (\ref{pot2}) is readily recognized as a complex, energy-dependent version of the Morse potential 
\cite{dere}. In contrast to the previous application, the present potential (\ref{pot2}) does not feature ${\cal PT}$-symmetry unless 
we make the trivial assignment $B=0$. Despite this fact, we can show that (\ref{pot2}) is pseudo-hermitian. To this end, 
we will use a result obtained in \cite{ahmed}. In the latter reference it was shown that pseudo-hermiticity is 
established by an operator $\eta$ of the form $\eta=\exp(-\theta~p)$, where $p=-i ~d/dx$ and $\theta$ is a real constant that we will 
determine below for the present system. Using the latter operator $\eta$, the relation 
$\eta V_E \eta^{-1}= V^\ast_E$ is equivalent to 
\begin{eqnarray}
V_E(x+i~\theta) &=& V_E^\ast(x), \nonumber
\end{eqnarray}
implying that the Hamiltonian associated with the system is $\eta$-pseudo-hermitian with respect to $\eta = \exp(-\theta p)$ 
(for the sake of brevity we will refer to our system simply as pseudo-hermitian). Including energy-dependence in the parameter 
$B$ allows to maintain pseudo-hermiticity, as we will demonstrate now. To this end, 
let us define a real-valued quantity $\theta$ by means of
\begin{eqnarray}
\theta &=& \frac{2}{\alpha}~\arctan\left\{\frac{\mbox{Im}[B(E)]}{\mbox{Re}[B(E)]}
\right\}. \nonumber
\end{eqnarray}
We can express $\theta$ through complex logarithms as follows
\begin{eqnarray}
\theta &=& \frac{i}{\alpha}~\log\left\{1-i~\frac{\mbox{Im}[B(E)]}{\mbox{Re}[B(E)]}
\right\}-\frac{i}{\alpha}~\log\left\{1+i~\frac{\mbox{Im}[B(E)]}{\mbox{Re}[B(E)]}
\right\}. \nonumber
\end{eqnarray}
We will now use this $\theta$ to verify pseudo-hermiticity of our potential (\ref{pot2}). First, we note that
\begin{eqnarray}
\exp\left[-\alpha~(x+i~\theta)\right] &=& \exp(-\alpha~x)~\exp(-i~\alpha~\theta) \nonumber \\[1ex]
&=&  \exp(-\alpha~x)~\left\{1-i~\frac{\mbox{Im}[B(E)]}{\mbox{Re}[B(E)]}
\right\}\left\{1+i~\frac{\mbox{Im}[B(E)]}{\mbox{Re}[B(E)]}
\right\}^{-1} \nonumber \\[1ex]
&=&  \exp(-\alpha~x)~\frac{B(E)^\ast}{B(E)}. \nonumber
\end{eqnarray}
Upon plugging this identity into our potential (\ref{pot2}), we obtain
\begin{eqnarray}
V_E(x+i~\theta) &=& A^2+\left[B(E)^\ast\right]^2 \exp\left(-2~\alpha~x \right)-2~B(E)^\ast \left(A+\frac{\alpha}{2} \right)
~\exp\left(-\alpha~x \right) ~=~ V_E^\ast(x). \nonumber
\end{eqnarray}
This relation establishes pseudo-hermiticity of (\ref{pot2}). Now, solutions to our boundary-value problem 
(\ref{bvp1}), (\ref{bvp2}) for the potential (\ref{pot2}) exist if the energy takes the following discrete values $E=E_n$, 
where
\begin{eqnarray}
E_n &=& A^2-\left(A-\alpha~n \right)^2, \label{ene2}
\end{eqnarray}
introducing a nonnegative integer $n$ satisfying $0 \leq n < A/\alpha$. As in the previously discussed case of the 
hyperbolic Scarf potential, we see that the energies (\ref{ene2}) are real despite the potential (\ref{pot2}) 
taking complex values. The solutions $\Psi=\Psi_n$, $0 \leq n < A/\alpha$, of our equation (\ref{bvp1}) and 
associated with our energies 
(\ref{ene2}) read 
\begin{eqnarray}
\Psi_n(x) &=& \exp\left[-A~x+\alpha~ n~ x-\frac{B(E_n)}{\alpha}~\exp(-\alpha~x)\right]~ L_n^{\frac{2A}{\alpha}-2n}
\left[\frac{2 B(E_n)}{\alpha}~\exp\left(-\alpha~ x\right)\right] \hspace{-.1cm} , \label{sol2}
\end{eqnarray}
where $L_n$ stands for an associated Laguerre polynomial of degree $n$ \cite{abram}. In order to verify that the functions (\ref{sol2}) 
satisfy the boundary conditions (\ref{bvp2}) at positive and negative infinity, let us inspect the first exponential factor on the 
right side of (\ref{sol2}). Due to the constraint $n < A / \alpha$, the latter factor, together with the Laguerre polynomial, 
dominates at positive infinity, guaranteeing exponential decay to zero in both real and imaginary part of the solutions. 
Next, the behaviour of (\ref{sol2}) at negative infinity is governed by the double exponential. While the imaginary part of 
$B$ produces an oscillatory function, its real part must be positive in order to ensure vanishing of the solutions at 
negative infinity. In summary, our boundary conditions (\ref{bvp2}) are fulfilled only if the real part of $B$ is 
positive. The remaining task is to evaluate orthogonality condition and pseudo-norm, which we will perform within the 
following example.

\begin{figure}[h]
\begin{center} \vspace{.3cm}

\epsfig{file=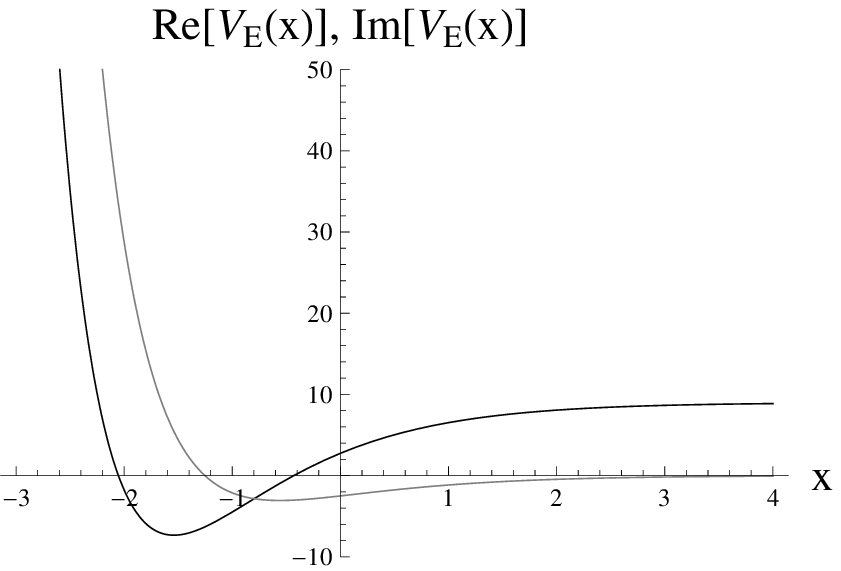,width=7.8cm}
\epsfig{file=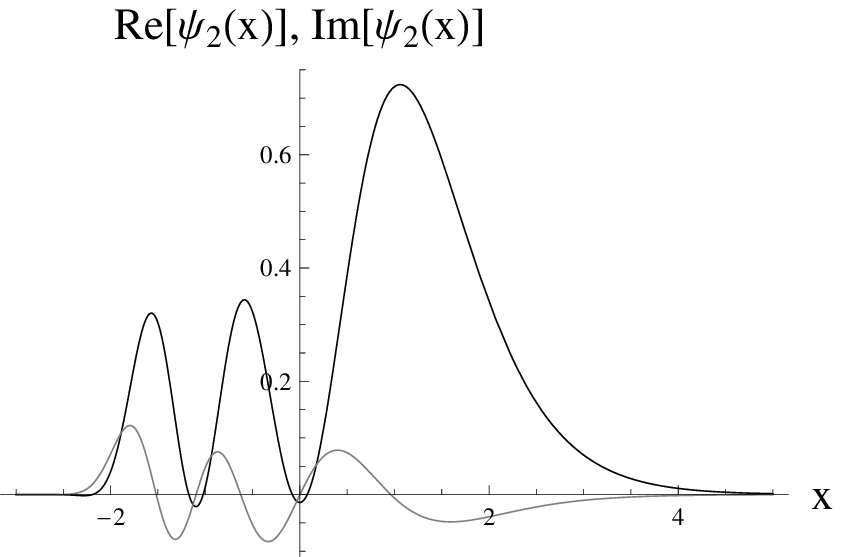,width=7.8cm}
\caption{Left plot: graph of the real part (black curve) and imaginary part (grey curve) of the potential (\ref{pot2}) 
with $n=1$ and $k=0.1$. Right plot: graph of the real part (black curve) and imaginary part (grey curve) of (\ref{dens}) for 
$n=2$ and $k=0.01$. }
\label{fig2}
\end{center}
\end{figure}

\paragraph{Example.} In order to evaluate the orthogonality condition and the pseudo-norm we now 
choose the parameters $A, \alpha$ and $B$ as follows
\begin{eqnarray}
A~=~3 \qquad \qquad \qquad \alpha~=~1 \qquad \qquad \qquad B(E)~=~1+i~k~E, \label{set2}
\end{eqnarray}
where $k>0$ is a constant. Recall that the real part of $B$ is allowed to depend on $E$. However, we chose it as a constant in order to 
simplify and shorten subsequent calculations. Similar to (\ref{set1}) and (\ref{set3}), parameter values were 
assigned in (\ref{set2}) such as to facilitate subsequent calculations. We now substitute the settings (\ref{set2}) into our potential (\ref{pot2}), 
rendering it in the form
\begin{eqnarray}
V_E(x) &=& 9-7~\exp(-x)~\left\{1+i~k~\left[9-(3-n)^2 \right] \right\}+\exp(-2~x) \times \nonumber \\[1ex]
&\times&
\left\{1+i~k~\left[9-(3-n)^2\right]\right\}^2. \label{pot2x}
\end{eqnarray}
Observe that this potential is real-valued for the particular value $n=0$. A particular example of (\ref{pot2x}) 
is displayed in figure \ref{fig2}. Next, we obtain the discrete energies by plugging (\ref{set2}) into (\ref{ene2}). This gives
\begin{eqnarray}
E_n &=& -n~(n-6),~~~n=0,1,2. \label{ene2x}
\end{eqnarray}
The solutions (\ref{sol2}) associated with the present settings (\ref{set2}) take the form
\begin{eqnarray}
\Psi_n(x) &=& \exp\Big\{-\exp(-x) \left[1+i~k~(6n-n^2) \right]+(n-3)~x\Big\} \times \nonumber \\[1ex]
&\times& L_n^{-2n+6}\left\{
2~ \exp(-x )\left[1+i~k~(6n-n^2) \right]
\right\}, \label{sol2x}
\end{eqnarray}
where $n=0,1,2$. Observe that the functions in (\ref{sol2x}) satisfy our boundary conditions (\ref{bvp2}) at the 
infinities because the real part of $B$ is positive, see (\ref{set2}). Our next task is to study orthogonality of the functions 
(\ref{sol2x}). As in the previous case of the 
hyperbolic Scarf potential, we will use a Wronskian representation, but this time we need the version (\ref{orthow}). 
For admissible integers $m,n$ 
satisfying $m \neq n$ we find the Wronskian of $\Psi^\ast_m$ and $\eta \Psi_n$ as
\begin{eqnarray}
W_{\Psi^\ast_m, \eta \Psi_n}(x) &=& \exp\Bigg\{\exp(-x) \Big\{
-2-i~k~\Big[m~(m-6)+n~(n-6) \Big]+ \nonumber \\[1ex]
& & \hspace{-2cm} +~\exp(x)~x~(m+n-7)-2~i~\exp(x)~(n-3)~\arctan\Big[k~n~(n-6) \Big]
\Big\}
\Bigg\} ~\times \nonumber \\[1ex]
& & \hspace{-2cm} \times~
\Bigg\{
-2~\Big[1+i~k~m~(m-6) \Big] L^\ast_{m-1}(x)~\hat{L}_n(x)+\Big[1+\exp(x)~(m-3)+ \nonumber \\[1ex]
& & \hspace{-2cm} +~i~k~m~(m-6)\Big]~L^\ast_m(x)~\hat{L}_n(x)+
L^\ast_m(x) \Bigg[2~\Big\{2~\exp\Big[2~i~\arctan[k~n~(n-6)]\Big] \times \nonumber \\[1ex]
& & \hspace{-2cm} \times~ \Big[1-i~k~n~(n-6)\Big]~L_{n-1}(x)+
\Big\{\exp(x)~(n-3)+\exp\Big[2~i~\arctan[k~n~(n-6)]\Big]\times \nonumber \\[1ex]
& & \hspace{-2cm} \times~
\Big[1-i~k~n~(n-6)\Big] [1-i~k~n~(n-6)]~\hat{L}_n(x)
\Big\}
\Bigg]
\Bigg\}
. \label{wron2}
\end{eqnarray}
For the sake of brevity we used the following abbreviations for the associated Laguerre polynomials that appear in 
the latter Wronskian
\begin{eqnarray}
L_{j-l}(x) \hspace{-.2cm}  &=& \hspace{-.2cm} \left\{ \begin{array}{llllll}
L_{j-l}^{l-2j+6}\left\{2 \exp(-x )\left[1-i~k~(6j-j^2) \right]\right\}
 & \mbox{if} & j-l \geq 0 \\[1ex]
0 & \mbox{if} & j-l<0
\end{array}
\right\} \nonumber \\[1ex]
\hat{L}_{j-l}(x) \hspace{-.2cm}  &=& \hspace{-.2cm} \left\{ \begin{array}{llllll}
L_{j-l}^{l-2j+6}\left\{2 \exp\Big[2~i~\arctan[k~n~(n-6)]\Big] \left[1-i~k~(6j-j^2) \right]\right\}
 & \mbox{if} & j-l \geq 0 \\[1ex]
0 & \mbox{if} & j-l<0
\end{array}
\right\}, \nonumber \\[1ex] 
\label{abbl}
\end{eqnarray}
where $j,l=0,1,2$. In order to show that the functions in (\ref{sol2x}) satisfy the orthogonality relation (\ref{orthow}), we 
study the behaviour of the Wronskian (\ref{wron2}) at the infinities. First we observe that both the first 
exponential factor on the right side of (\ref{wron2}) as well as the Laguerre polynomials tend to zero in real and 
imaginary part as $x$ goes to positive infinity. As a consequence, the Wronskian vanishes there. The behaviour of 
our Wronskian at negative infinity is governed by terms located in the first exponential factor on the 
right side of (\ref{wron2}). More precisely, the terms dominant at negative infinity read
\begin{eqnarray}
W_{\Psi^\ast_m,\eta \Psi_n}(x) &\stackrel{x << -1}{\sim}  
 \exp\Bigg\{-\exp(-x)~ \Big\{
2+i~k~\Big[m~(m-6)+n~(n-6) \Big]\Big\}\Bigg\}. \label{dom}
\end{eqnarray}
This term vanishes at negative infinity in both real and imaginary part. Even though the Laguerre polynomials 
in (\ref{wron2}) grow exponentially as $x$ tends to negative infinity, they 
cannot compensate the double exponential decay from (\ref{dom}). Consequently, the Wronskian (\ref{wron2}) 
vanishes at both infinities. According to (\ref{orthow}), this implies orthogonality of the functions in (\ref{sol2x}). 
The final task is now to prove normalizability of the latter functions. Since there are only three of such functions 
due to $n<3$, we proceed in a way that is similar to the previous application. In the first step we establish existence 
of the integral in (\ref{norm}). To this end, we evaluate
\begin{eqnarray}
\Psi^\ast_n(x)~\eta~\Psi_n(x) \left[1-\frac{\partial V^\ast_{E}(x)}{\partial E}\right]_{\big| E=E_n} \hspace{-.4cm} &=& \nonumber \\[1ex]
& & \hspace{-5cm}=~\exp\Bigg\{
-\exp(-x)~\Big\{-2-2~i~k~n~(n-6)+2~\exp(x)~x~(n-4)- \nonumber \\[1ex]
& & \hspace{-5cm}
-~2~i~\exp(x)~(n-3)~\arctan[k~n~(n-6)]\Big\}
\Bigg\}
\Bigg\{
\exp(2~x)-7~i~k~\exp(x)+ \nonumber \\[1ex]
& & \hspace{-5cm}+~
2~k~[i-k~n~(n-6)]
\Bigg\}
~L^\ast_n(x)~\hat{L}_n(x)
. \label{dens} 
\end{eqnarray}
Note that here we have used once more the abbreviations (\ref{abbl}) for the associated Laguerre polynomials. The right plot 
in figure \ref{fig2} shows an example of (\ref{dens}).
It turns out that the function (\ref{dens}) vanishes at the infinities. The reasoning is very similar to the argumentation 
that we presented above for the case of our Wronskian (\ref{wron2}), such that we omit to show it here. It follows that 
the integral in (\ref{norm}) exists in the present case, such that we can proceed to calculate its value numerically. 
We find for the three applicable cases
\begin{eqnarray}
N(\Psi_0) &=& 1.875 \nonumber \\[1ex] 
N(\Psi_1) &=& 1.86566 \nonumber \\[1ex]
N(\Psi_2) &=& 1.49046.  \label{densx}
\end{eqnarray}
The correctly normalized solutions (\ref{sol2x}) are now obtained from (\ref{densx}) by means of 
$\Psi_n / N(\Psi_n)$, $n=0,1,2$.

\section{Concluding remarks}
In this work we have studied examples of quantum systems featuring complex-valued, energy-dependent potentials 
that admit real energy spectra and normalizable solutions forming orthogonal sets. While bound-state solutions belonging to 
pseudo-hermitian or ${\cal PT}$-symmetric systems are usually normalized in a weighted $L^2$-space, in the present 
cases we have normalized using the definition (\ref{norm}), which stems from the modified quantum theory applicable 
to systems with energy-dependent potentials. Investigation of the relationship between both types of normalization is 
subject to future research. Before we conclude this work, let us comment on the applicability of the results that we have 
obtained. One of the principal application fields of complex potentials featuring energy-dependence is nuclear physics 
\cite{alberi} \cite{mckellar} \cite{miya}, particularly the calculation of expectation values and bound-state normalization 
constants. Since for energy-dependent potentials one has to adhoc-modify the inner product and norm of the former underlying 
Hilbert space, there are several possibilities of doing so. Besides the conventional approach \cite {hokkyo} 
\cite{hokkyoerr} \cite{miya}, for pseudo-hermitian or ${\cal PT}$-symmetric potentials our work suggests 
alternative ways of calculating expectation values and determining normalization constants through our 
modified norms (\ref{norm}) and (\ref{normpt}), respectively. In practical applications such as \cite{miya}, 
results given by our norms may be compared with existing results and experimentally obtained findings.

\end{sloppypar}

\end{document}